\PassOptionsToPackage{unicode}{hyperref}
\PassOptionsToPackage{hyphens}{url}
\PassOptionsToPackage{dvipsnames,svgnames,x11names}{xcolor}
\documentclass[
  journal,
]{IEEEtran}%
\usepackage[cmex10]{amsmath}
\usepackage{amssymb}
\usepackage{iftex}
\ifPDFTeX
  \usepackage[T1]{fontenc}
  \usepackage[utf8]{inputenc}
  \usepackage{textcomp} 
\else 
  \usepackage{unicode-math} 
  \defaultfontfeatures{Scale=MatchLowercase}
  \defaultfontfeatures[\rmfamily]{Ligatures=TeX,Scale=1}
\fi
\ifPDFTeX\else
\fi
\IfFileExists{upquote.sty}{\usepackage{upquote}}{}
\IfFileExists{microtype.sty}{
  \usepackage[]{microtype}
  \UseMicrotypeSet[protrusion]{basicmath} 
}{}
\makeatletter
\ifCLASSOPTIONcompsoc
\fi
\makeatother
\usepackage{xcolor}
\ifx\paragraph\undefined\else
  \let\oldparagraph\paragraph
  \renewcommand{\paragraph}[1]{\oldparagraph{#1}\mbox{}}
\fi
\ifx\subparagraph\undefined\else
  \let\oldsubparagraph\subparagraph
  \renewcommand{\subparagraph}[1]{\oldsubparagraph{#1}\mbox{}}
\fi

\usepackage{longtable,booktabs,array}
\usepackage{calc} 
\usepackage{etoolbox}
\makeatletter
\patchcmd\longtable{\par}{\if@noskipsec\mbox{}\fi\par}{}{}
\makeatother
\IfFileExists{footnotehyper.sty}{\usepackage{footnotehyper}}{\usepackage{footnote}}
\makesavenoteenv{longtable}
\usepackage{graphicx}
\makeatletter
\newsavebox\pandoc@box
\newcommand*\pandocbounded[1]{
  \sbox\pandoc@box{#1}%
  \Gscale@div\@tempa{\textheight}{\dimexpr\ht\pandoc@box+\dp\pandoc@box\relax}%
  \Gscale@div\@tempb{\linewidth}{\wd\pandoc@box}%
  \ifdim\@tempb\p@<\@tempa\p@\let\@tempa\@tempb\fi
  \ifdim\@tempa\p@<\p@\scalebox{\@tempa}{\usebox\pandoc@box}%
  \else\usebox{\pandoc@box}%
  \fi%
}
\def\fps@figure{htbp}
\makeatother

\NewDocumentCommand\citeproctext{}{}
\NewDocumentCommand\citeproc{mm}{%
  \begingroup\def\citeproctext{#2}\cite{#1}\endgroup}
\makeatletter
 \let\@cite@ofmt\@firstofone
 \def\@biblabel#1{}
 \def\@cite#1#2{{#1\if@tempswa , #2\fi}}
\makeatother
\newlength{\cslhangindent}
\newlength{\csllabelwidth}
\newenvironment{CSLReferences}[2] 
 {\begin{list}{}{%
  \setlength{\itemindent}{0pt}
  \setlength{\leftmargin}{0pt}
  \setlength{\parsep}{0pt}
  \ifodd #1
   \setlength{\leftmargin}{\cslhangindent}
   \setlength{\itemindent}{-1\cslhangindent}
  \fi
  \setlength{\itemsep}{#2\baselineskip}}}
 {\end{list}}
\usepackage{calc}

\newcommand{\CSLLeftMargin}[1]{\parbox[t]{\csllabelwidth}{\strut#1\strut}}
\newcommand{\CSLRightInline}[1]{\parbox[t]{\linewidth - \csllabelwidth}{\strut#1\strut}}

\providecommand{\tightlist}{%
  \setlength{\itemsep}{0pt}\setlength{\parskip}{0pt}}

\usepackage[ruled,vlined,commentsnumbered]{algorithm2e}
\usepackage{physics}
\usepackage[version=3]{mhchem}
\usepackage{orcidlink}
\usepackage{float}
\makeatletter
\@ifpackageloaded{caption}{}{\usepackage{caption}}
\AtBeginDocument{%
\ifdefined\contentsname
  \renewcommand*\contentsname{Table of contents}
\else
  \newcommand\contentsname{Table of contents}
\fi
\ifdefined\listfigurename
  \renewcommand*\listfigurename{List of Figures}
\else
  \newcommand\listfigurename{List of Figures}
\fi
\ifdefined\listtablename
  \renewcommand*\listtablename{List of Tables}
\else
  \newcommand\listtablename{List of Tables}
\fi
\ifdefined\figurename
  \renewcommand*\figurename{Fig.}
\else
  \newcommand\figurename{Fig.}
\fi
\ifdefined\tablename
  \renewcommand*\tablename{Table}
\else
  \newcommand\tablename{Table}
\fi
}
\@ifpackageloaded{float}{}{\usepackage{float}}
\floatstyle{ruled}
\@ifundefined{c@chapter}{\newfloat{codelisting}{h}{lop}}{\newfloat{codelisting}{h}{lop}[chapter]}
\floatname{codelisting}{Listing}

\usepackage{amsthm}
\theoremstyle{plain}
\newtheorem{lemma}{Lemma}[section]
\theoremstyle{remark}
\AtBeginDocument{}

\makeatother
\makeatletter
\@ifpackageloaded{caption}{}{\usepackage{caption}}
\@ifpackageloaded{subcaption}{}{\usepackage{subcaption}}
\makeatother
\usepackage[skip=2pt,font=footnotesize]{caption}
\makeatletter
\providecommand{\bibfont}{\footnotesize}
\let\CSLReferences@rig=\CSLReferences
\renewcommand{\CSLReferences}[2]{
\bibfont\settowidth\csllabelwidth{[999]}
\CSLReferences@rig{#1}{#2}
\vskip 0.3\baselineskip plus 0.1\baselineskip minus 0.1\baselineskip%
}
\makeatother
\ifLuaTeX
  \usepackage{selnolig}  
\fi
\IfFileExists{bookmark.sty}{\usepackage{bookmark}}{\usepackage{hyperref}}
\IfFileExists{xurl.sty}{\usepackage{xurl}}{} 
\hypersetup{
  pdftitle={Anchored Iterative Decoding of Reed--Solomon Product Codes for High-Rate Magnetic Tape Storage},
  pdfauthor={Alexander Chvanov; Nikolay Malkovsky},
  pdfkeywords={product codes, Reed--Solomon codes, magnetic tape
storage, anchor decoding, iterative decoding, error floor},
  colorlinks=true,
  linkcolor={blue},
  filecolor={Maroon},
  citecolor={Blue},
  urlcolor={Blue},
  pdfcreator={LaTeX via pandoc}}

\title{Anchored Iterative Decoding of Reed--Solomon Product Codes for
High-Rate Magnetic Tape Storage}

\author{
\parbox{0.45\textwidth}{\centering Alexander
Chvanov\\[0.3ex]{\small\itshape Chebyshev Research
Center}\\[0.2ex]{\small\texttt{505995@niuitmo.ru}}}\hspace{1em}\parbox{0.45\textwidth}{\centering Nikolay
Malkovsky\\[0.3ex]{\small\itshape Chebyshev Research
Center}\\[0.2ex]{\small\texttt{malkovskynv@gmail.com}}}%
}
\begin{document}



\maketitle

\begin{abstract}
We present a decoding algorithm for a Reed--Solomon product code
targeted at the demanding reliability requirements of modern magnetic
tape storage. For a code \(RS(256, 224)\times RS(175, 173)\) in a BSC we
achieve post-decoding bit error rate (BER) of \(10^{-20}\) at
\(\approx 5.3\times 10^{-3}\) raw BER while maintaining the common
requirement to detect and correct \(4\) out of \(32\) dead tracks. We
resolve two common problems of miscorrections and stall patterns that
arise under such a high raw BER via simplified versions of anchor
decoding and binary image analysis. The highest efficiency is achieved
by performing an unlimited iterative decoder, which performs decoding
iterations until no changes are made. We give an empirical analysis of
the average number of iterations: below the raw BER of
\(5\times 10^{-3}\), four half-iterations are performed on average while
improving substantially over a decoder that always performs \(4\)
half-iterations. We provide a CPU implementation achieving 100-200
megabytes/s single-core throughput on a regular laptop and a design of a
potential hybrid FPGA-CPU implementation of our decoder.
\end{abstract}

%

\section{Introduction}\label{sec-intro}

\emph{Product codes} are a family of error correction codes (ECC)
introduced by Elias \citeproc{ref-elias1954_product}{{[}1{]}}. They are
constructed by two or more components each one itself being an ECC. In
particular, a two-dimensional \(A\times B\) product code is formed by
first organizing the data into a table of symbols, adding parity symbols
to every row by code \(B\) and then adding parity for every column
(including columns introduced by \(B\)) via code \(A\). The construction
can be viewed as a special case of \emph{concatenated codes}
\citeproc{ref-forney1966concatenated}{{[}2{]}}; the baseline decoding
procedure of such a code is to first perform decoding on columns
followed by decoding on rows. In particular, for product codes with
bounded distance decoding some columns might become decodable after we
decode rows. This leads to an \emph{iterative bounded-distance
decoding}: a procedure where we iterate between column and row decoding
until some criterion is met, usually either the iteration limit is
reached or an iteration introduces no changes. In some sense this is
structurally similar to belief propagation decoders while being a
hard-decision decoder. This kind of procedure is explored in classical
analyses of product-code performance under iterated decoding
\citeproc{ref-justesen2011_product}{{[}3{]}} and remains an active
research direction, with recent work on miscorrection-aware decoding
\citeproc{ref-hager_pfister2018_anchor}{{[}4{]}},
\citeproc{ref-peng2024_grand_anchor}{{[}5{]}} and error-floor analysis
\citeproc{ref-holzbaur2019_staircase}{{[}6{]}},
\citeproc{ref-hager2016_density_evolution_isit}{{[}7{]}}.

Product codes and iterative decoders have found widespread use in
optical transport networks \citeproc{ref-smith2012_staircase}{{[}8{]}}
and magnetic tape storage
\citeproc{ref-furrer2021317}{{[}9{]}}--\citeproc{ref-arslan2022_lto9_uber}{{[}12{]}}.
In HDD and SSD storage, by contrast, LDPC codes --- often concatenated
with an outer RS code --- are the dominant ECC scheme
\citeproc{ref-hareedy2016_ldpc_hdd}{{[}13{]}}--\citeproc{ref-zhao2013_ldpc_ssd}{{[}15{]}}.

Modern magnetic tape storage poses the following reliability challenges:

\begin{itemize}
\tightlist
\item
  Individual bit flips
\item
  Track malfunctions
\item
  Other physical disruptions
\end{itemize}

The last two issues are common in the sense that both produce burst
errors. A common practical solution to general burst errors is
interleaving data of several blocks to spread the errors between them;
this is also commonly used in magnetic tapes. For track malfunctions,
interleaving does not help; this is where product Reed--Solomon codes
have particular interest as they provide additional functionality for
track malfunctions: to compensate for random bit errors, the code is
usually decoded with a fixed number of iterations to achieve the desired
post-decoding BER and stable throughput; to compensate for dead tracks,
the rows are usually placed on a single track and are used as detectors
and pass information to the column decoder to be used in erasure or
error-plus-erasure modes.

In this paper we present a few practical algorithm improvements over a
standard iterative bounded distance decoding of product RS codes that
are designed for use in the magnetic tape scenario. In particular, for
the code \(RS(256, 224)\times RS(175, 173)\) the following is achieved:

\begin{itemize}
\tightlist
\item
  Recovery of \(4\) out of \(32\) dead tracks via the inner code, by
  design aligning with the setup in
  \citeproc{ref-furrer2021317}{{[}9{]}},
  \citeproc{ref-lantz2025_tape_storage}{{[}10{]}}.
\item
  Very high code rate of \(86.50\%\) compared to \(\sim 81-83\%\) for
  LTO-9 and prior standards.
\item
  Post-decoding BER of \(10^{-20}\) is achieved at
  \(\sim 5.3\times 10^{-3}\) raw BER under a binary symmetric channel.
\item
  Ability to detect dead tracks with probability \(1-10^{-20}\) using
  \(90\) megabytes of data under a raw BER of \(5\times 10^{-3}\) and
  \(18\) kilobytes of data under a raw BER of \(5\times 10^{-4}\).
\end{itemize}

To achieve provable post-decoding BER at high code rates we address
miscorrections and stall patterns and use an unlimited iterative
decoder. Our proof is not complete but follows common simulation
principles supported by edge case analysis of miscorrections and stall
patterns. The use of an unlimited iterative decoder is not practical in
general but we show that most of the cases are covered by \(3\)
half-iterations (column-row-column) and, moreover, decoding can be
organized in a way that only the first two half-iterations are
computationally intensive.

The remainder of the paper is organized as follows.
Section~\ref{sec-background} reviews the ECC setup, the LTO-9 product
code construction, and the related problems of miscorrections and stall
patterns. Section~\ref{sec-construction} describes dead-track detection,
the modified decoding algorithm, and its analysis.
Section~\ref{sec-evaluations} presents the empirical results, and
Section~\ref{sec-conclusion} concludes with practical considerations.

\section{Background and related work}\label{sec-background}

\subsection{Error correction codes in
LTO}\label{error-correction-codes-in-lto}

The main ingredient of the ECC for LTO is the Reed--Solomon (RS) code
\citeproc{ref-rs_origin_paper}{{[}16{]}}. Code \(RS(n, k)\) over a field
\(\mathbb{F}\) encodes arbitrary \(k\)-length sequence
\(\mathbf{x}=(x_1, \ldots, x_k)\in \mathbb{F}^k\) into a unique
\(n\)-length sequence \(\mathbf{y}=(y_1, \ldots, y_n)\in \mathbb{F}^n\)
forming an encoding map \(Enc: \mathbb{F}^k\rightarrow \mathbb{F}^n\);
elements of the image \(\Im (Enc)\) are usually referred to as
\emph{codewords}. The most important property of the code is that for
distinct \(\mathbf{x}_1\neq \mathbf{x}_2\) \[
Hamming(Enc(\mathbf{x}_1), Enc(\mathbf{x}_2))\geq n-k+1=d,
\] where \(Hamming(\mathbf{y}, \mathbf{z})=|\{y_i\neq z_i\}|\) --- the
standard Hamming distance. From a practical point of view, this means
that the code can

\begin{itemize}
\tightlist
\item
  Detect and correct up to \(\lfloor(d-1)/2\rfloor\) errors in
  \textbf{error correction mode}
\item
  Detect but not correct up to \(d-1\) errors in \textbf{error detection
  mode}
\item
  Correct up to \(d-1\) errors if their positions are known, such errors
  are commonly referred to as \emph{erasures} and we call such decoding
  \textbf{erasure correction mode}
\item
  Detect and correct up to \(t\) errors when \(e\) erasures are present
  and \(2t+e<d\), such decoding is \textbf{erasure and error correction
  mode}
\end{itemize}

RS codes are among the most widely used types of codes and have been
extensively explored since their discovery
\citeproc{ref-rs_origin_paper}{{[}16{]}}--\citeproc{ref-roth2006_coding_theory}{{[}22{]}}.
A \emph{systematic} RS code is an RS code with the additional property
that \(Enc(\mathbf{x})_i=x_i, i\leq k\), i.e.~the encoding procedure
adds parity to the original data rather than completely transforming it.
If \(RS(n, k)\) exists then its systematic variant also exists; for the
rest of the paper we will operate with systematic codes.

\begin{figure}

\centering{

\pandocbounded{\includegraphics[keepaspectratio]{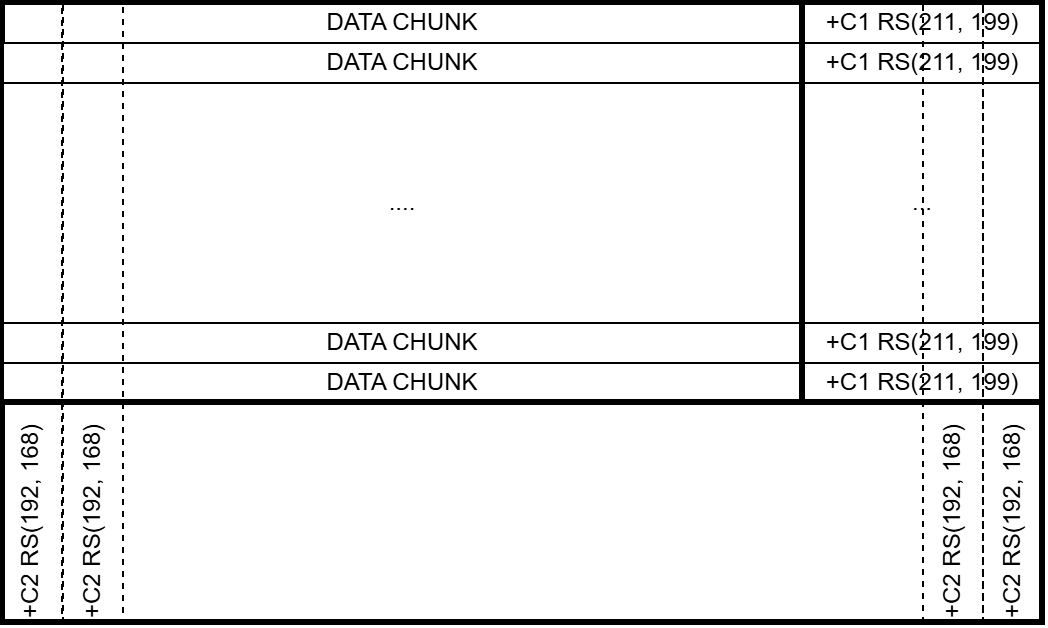}}

}

\caption{\label{fig-product-code}An example of an
\(RS(192, 168)\times RS(211, 199)\) product code.}

\end{figure}%

For an RS product code \(RS(n_1, k_1)\times RS(n_2, k_2)\) a single
block is formed by \(k_1\times k_2\) field elements organized in a
table. Encoding first consists of appending \(n_2-k_2\) parity elements
to each of \(k_1\) rows forming row codewords followed by appending
\(n_1-k_1\) elements to each of the \(n_2\) columns. There is an
important non-obvious property of this construction, although by
construction we get \(k_1\) rows forming codewords, the remaining
\(n_1-k_1\) rows are formed by a column encoding. Surprisingly, these
\(n_1-k_1\) bottom rows also form valid codewords; this is implied by
the linearity of the RS code.

The IBDD of a product code is then the following procedure: alternate
between column and row iterations, on row iterations apply decoding in
error correction mode for every row, on column iterations --- for every
column; stop when the iteration limit is reached or an iteration
produces no changes.

\begin{algorithm}[t]
\SetKwFunction{Decode}{Decode}
\SetKwFunction{Changed}{Changed}
\KwData{Product code block $E\in \mathbb{F}_{q}^{n_1\times n_2}$ and an iteration limit $I_{\max}$. $\Decode$ is the standard RS decoder in error mode, i.e. for an $RS(n, k)$ code, find and correct up to $\lfloor(n-k)/2\rfloor$ errors; it returns $\textsc{true}$ if a correction was made.}
\KwResult{In-place error correction of $E$.}
\For{$t=1, 2, \ldots, I_{\max}$}{
  $c \leftarrow \textsc{false}$ \tcp*{change flag}
  \If(\tcp*[f]{column iteration}){$t$ is odd}{
    \For{$j=1, \ldots, n_2$}{
      $\tilde{c} \leftarrow \Decode(E[:,j])$\;
      \If{$\tilde{c}$}{
        $c \leftarrow \textsc{true}$\;
      }
    }
  }
  \Else(\tcp*[f]{row iteration}){
    \For{$i=1, \ldots, n_1$}{
      $\tilde{c} \leftarrow \Decode(E[i,:])$\;
      \If{$\tilde{c}$}{
        $c \leftarrow \textsc{true}$\;
      }
    }
  }
  \If{not $c$}{
    \Return \;
  }
}
\caption{Iterative bounded distance decoder (IBDD) for an
$RS$ product code.}
\label{alg:ibdd}
\end{algorithm}

Notice that in both row and column iterations, all rows/columns are
decoded relying on the linearity property previously discussed.

\subsection{RS product codes setup for
LTO-9}\label{rs-product-codes-setup-for-lto-9}

According to \citeproc{ref-arslan2022_lto9_uber}{{[}12{]}}, a ``dead
track'' error is a problem that happens when we write the data and we
want to correct it via a secondary module with no access to the initial
data. To address the issue of dead tracks via RS product codes, row
codewords are spread across the tracks; each individual codeword is
always placed onto a single track. With this, analyzing several row
codewords from a particular track allows us to detect whether it is dead
or alive. When a dead track is detected, the information is passed to
the column decoder; then \autoref{alg:ibdd} is employed with the column
decoder working in error and erasure mode with rows corresponding to
dead tracks being marked as erasures. In an extreme case, when the
number of dead tracks reaches the column code capability, the row
decoder is required to correct bit errors by itself. When on-write
validation and correction succeeds, the tape should then be readable for
a long period of time, while some gradual degradation is expected.

For LTO-9, \(32\) tracks are used, and corruption of \(4\) tracks should
be tolerated. A code with parameters
\(RS(32\ell, 28\ell)\times RS(n, k)\) is usually used; \(\ell\) row
codewords are then placed on each track and recovery from \(4\) dead
tracks is guaranteed by design, assuming that the on-write BER is low
enough that \(RS(n, k)\) can correct them. In this paper we keep a
similar design but focus on the long-term degradation with high BER; a
complete analysis of the on-write correction is beyond the scope of this
paper.

\subsection{Miscorrections and stall
patterns}\label{miscorrections-and-stall-patterns}

While RS codes are quite easy to analyze, no complete theoretical
analysis of product RS codes exists. For example, the best attempts of
approximating the behavior of \autoref{alg:ibdd} are due to
\citeproc{ref-product_code_lower_bound}{{[}23{]}}; the approximation is
used in \citeproc{ref-furrer2021317}{{[}9{]}} and is reported to
underestimate for more than \(2\) iterations compared with simulations.
Monte-Carlo simulation is a more accurate way of approximating but is
limited to high raw BER; proving a post-decoding BER of \(10^{-20}\)
with a simulation will require simulating at least \(10^{20}\) random
bits, which is possible but expensive. A cheap variant is to estimate at
a high raw BER where the post-decoding BER is around \(10^{-10}\) and
extrapolate, but the approximation becomes inaccurate in this case.

Another issue with the simulation is that the post-decoding BER curve of
RS product codes is not convex but rather sigmoid-like; this is caused
by the \emph{stall patterns}; the effect is common for many ECC
families; for RS product codes, stall patterns are explored by
\citeproc{ref-hager_pfister2018_anchor}{{[}4{]}},
\citeproc{ref-product_codes_stall_patterns1}{{[}24{]}},
\citeproc{ref-product_codes_stall_patterns2}{{[}25{]}}.

\begin{figure}

\centering{

\pandocbounded{\includegraphics[keepaspectratio]{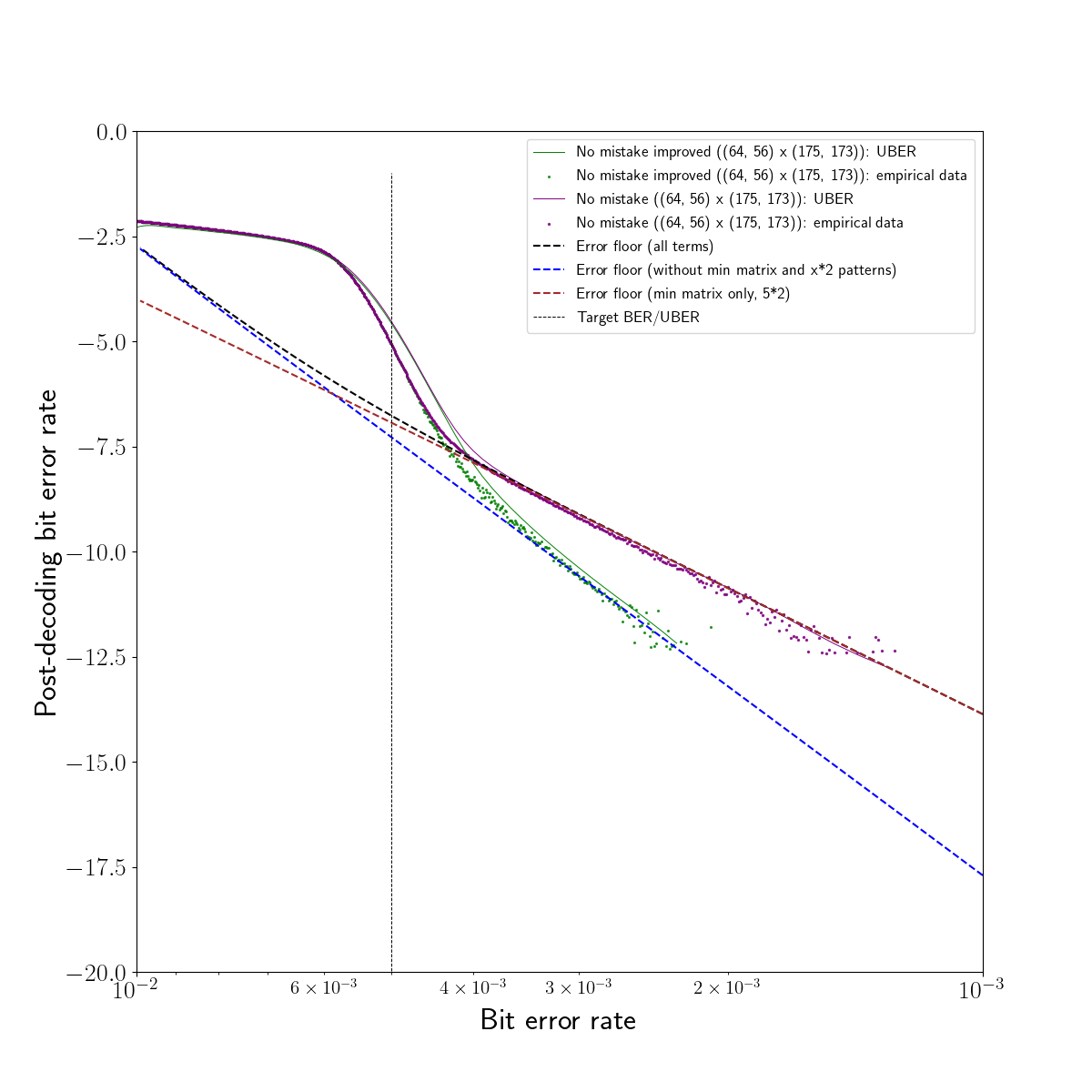}}

}

\caption{\label{fig-stall-patterns-monte-carlo}An example of stall
patterns for short codes. Two simulations with a miscorrection-free
decoder are provided: with and without trivial stall pattern recovery;
both simulations match the lower bounds given by
\citeproc{ref-product_codes_stall_patterns1}{{[}24{]}} and
\citeproc{ref-product_codes_stall_patterns2}{{[}25{]}}.}

\end{figure}%

Fig.~\ref{fig-stall-patterns-monte-carlo} gives a good reproducible
example of how stall patterns affect decoding. \autoref{alg:ibdd} will
get stuck on a trivial stall pattern and we will later show how to
resolve it.

Another common issue is \emph{miscorrection}: when the number of errors
exceeds the error correction capability of the RS code, there is a small
probability that this would be falsely detected as a correctable
codeword and corrected to the wrong codeword.
\citeproc{ref-hager_pfister2018_anchor}{{[}4{]}} gives a good algorithm
to detect miscorrections in RS product codes, but it is aimed at a
symmetric product code and performs an irregular backtracking procedure,
which is probably hard to implement on the hardware level. Still, the
ideas from their \emph{anchor} decoding form a strong basis for our own
algorithm.

\section{Main results}\label{sec-construction}

Our main results are for \(RS(256, 224)\times RS(175, 173)\) with a code
rate of \(86.50\%\).

\subsection{Dead track detection via high-rate RS
codes}\label{dead-track-detection-via-high-rate-rs-codes}

The first problem to be addressed is that dead track detection is much
harder with high-rate codes due to miscorrections. The standard approach
for dead track detection would be to check how many row codewords are
unrecoverable. For example, under a raw BER of \(10^{-3}\) an
\(RS(243, 231)\) code will have a \(3.7\times 10^{-3}\) failure rate and
a negligible miscorrection probability; having more than \(10\%\)
unrecoverable codewords would be a sign of a dead track with a
confidence level above \(1-10^{-20}\). For \(RS(175, 173)\), the
situation is different: recovery probability is \(\sim 0.41\) and the
miscorrection probability is of the same magnitude. With this, we cannot
rely on any kind of statistical test based on codeword recoverability.
Surprisingly, we can still detect dead tracks by checking the portion of
already correct codewords.

Here is a more detailed setup. We are trying to distinguish between two
situations:

\begin{itemize}
\tightlist
\item
  We are looking at a data corrupted by BSC with raw BER of \(p\)
\item
  We are looking at a completely random data of a dead track
\end{itemize}

The common approach is to perform some statistical test to distinguish
between these cases based on the sampled distribution of the data. We
assume that in the first case the data is initially a set of \(m\)
codewords of \(RS(n, k)\) over \(\mathbb{F}_{2^b}\); each bit is
corrupted independently with probability \(p\); for the second case the
data is \(m\) uniformly random \(n\)-length byte sequences. The
probabilities of a correct codeword are then

\begin{itemize}
\tightlist
\item
  For a BER of \(p\) the probability of no bit flips at all is
  \begin{equation}\phantomsection\label{eq-no-bit-errors}{
  (1-p)^{bn}.
  }\end{equation}
\item
  For a uniformly random byte sequence: for an arbitrary \(k\)-length
  sequence there is a unique \(n-k\)-byte parity that forms a valid
  codeword; thus, the probability of a correct codeword is
  \begin{equation}\phantomsection\label{eq-random-sequence-becomes-codeword}{
  \frac{1}{2^{b(n-k)}}.
  }\end{equation}
\end{itemize}

We additionally note that in the first case a correct codeword can be
formed by several bit errors, which increases the probability of getting
a correct codeword; we will ignore this probability to simplify the
approach without compromising it. For \(n=175, k=173\), \(b=8\), the
probabilities above meet at \(p\approx 0.008\) and thus in theory we can
distinguish for any \(p\) below that threshold via a Neyman-Pearson
test. In particular, for a raw BER of \(5\times 10^{-3}\),
(\ref{eq-no-bit-errors}) is \(8.9\times 10^{-4}\) and we need around
\(10^4-10^5\) codewords for a reliable test, resulting in several
megabytes of data; \(2^{19}\) codewords give a confidence level above
\(1-10^{-20}\), which amounts to \(\sim 90\) megabytes of data. Under a
raw BER of \(5\times 10^{-4}\) the probability (\ref{eq-no-bit-errors})
is \(\approx \frac{1}{2}\); \(6\) out of \(100\) correct codewords give
a confidence level above \(1-10^{-21}\).

\subsection{Modified IBDD algorithm}\label{modified-ibdd-algorithm}

IBDD (\autoref{alg:ibdd}) was investigated in
\citeproc{ref-elias1954_product}{{[}1{]}}--\citeproc{ref-holzbaur2019_staircase}{{[}6{]}},
\citeproc{ref-smith2012_staircase}{{[}8{]}},
\citeproc{ref-kschischang_sorokine1995_trellis}{{[}26{]}}. Our algorithm
is similar and is inspired by three core ideas:

\begin{itemize}
\tightlist
\item
  Analysis of the binary representation of RS codes
\item
  Anchor decoding of algebraic product codes
  \citeproc{ref-hager_pfister2018_anchor}{{[}4{]}}
\item
  Postprocessing in erasure mode
\end{itemize}

The LTO setup uses \(RS\) codes over the field \(\mathbb{F}_{2^8}\),
which naturally represents bytes. The errors, however, happen with
individual bits creating a domain mismatch. A natural way of
representing an \(\mathbb{F}_{2^8}\) field element is via a degree \(7\)
polynomial over \(\mathbb{F}_{2}\): coefficients of such a polynomial
form bits, and industrial hardware is typically tuned for this format.
This is, however, not a mandatory representation; for example, we might
want to use a non-standard basis for polynomials such as a Cantor basis
or a normal basis; in the general case, an arbitrary bijection
\(\varphi: \mathbb{F}_{2^8}\rightarrow \mathbb{F}_2^8\) can be used for
special properties. \(RS(n, k)\) is an MDS code meaning that its
distance is \(n-k+1\) in the field sense; the distance in the sense of
its binary representation might vary based on the exact encoding scheme
and the binary mapping in use. Many of these questions are explored in
the literature \citeproc{ref-macwilliams_sloane1977}{{[}21{]}},
\citeproc{ref-blake_kith1991_complete_weight}{{[}27{]}}--\citeproc{ref-elkhamy_mceliece2005_partition_weight}{{[}29{]}}
although few have practical use. We found a very simple heuristic based
on the binary representation, which suits our IBDD setup very well:
\emph{if \(RS(175, 173)\) corrects more than \(b=2\) bits in a byte,
this is most likely a miscorrection}. This is a simple technique that
allows us to reduce the number of miscorrections by the \(RS(175, 173)\)
decoder; it also false-detects some valid corrections, which is
tolerable since they can still be resolved by a column iteration.

\emph{Anchor decoding} \citeproc{ref-hager_pfister2018_anchor}{{[}4{]}}
is another technique developed to address miscorrections: when a
codeword is decoded incorrectly, it usually leads to conflicts with
other components; the idea is to disallow further corrections to bytes
already being corrected while backtracking conflicts between column and
row codewords and cancel decoding results when a single codeword takes
part in many conflicts. We adopted this idea for the LTO case in two
parts:

\begin{itemize}
\tightlist
\item
  The first part is that \(RS(256, 224)\) is much more reliable than
  \(RS(175, 173)\) and it is convenient to disallow the latter from
  changing after the former.
\item
  The second part is part of postprocessing to trace rare miscorrections
  of \(RS(256, 224)\): if at least \(33\) of \(RS(175, 173)\) codewords
  have syndromes and detect an error in the same column --- this is most
  likely a result of \(RS(256, 224)\) miscorrection.
\end{itemize}

\begin{algorithm}[t]
\SetKwFunction{FindErrors}{FindErrors}
\SetKwFunction{CorrectErrors}{CorrectErrors}
\SetKwFunction{And}{AND}
\SetKwFunction{False}{False}
\SetKwFunction{True}{True}
\SetKwFunction{HammingWeight}{HammingWeight}
\KwData{Encoded data $E\in \mathbb{F}_{2^8}^{n_r\times n_c}$ with $n_r$
rows and $n_c$ columns.}
\KwResult{In-place error correction of $E$.}
$C \leftarrow \{1, 2, \ldots, n_c\}$\;
$R \leftarrow \{1, 2, \ldots, n_r\}$\;
$C_{anchor} \leftarrow \emptyset$\;
\For{$t=1, 2, \ldots$ stop when $C\cup R=\emptyset$}{
  \If(\tcp*[h]{column iteration}){$t$ is odd}{
    \For{$j\in C$}{
      $E_c\leftarrow \FindErrors(E[:,j])$\;
      $R\leftarrow R\cup E_c$\;
      \If(\tcp*[h]{anchoring when no syndromes left}){$\CorrectErrors(E[:,j])$}{
        $C_{anchor}\leftarrow C_{anchor}\cup \{j\}$\;
      }
    }
    $C\leftarrow \emptyset$\;
  }
  \Else(\tcp*[h]{row iteration}){
    \For{$i\in R$}{
      $E_r\leftarrow \FindErrors(E[i,:])$\;
      \If{$E_r\cap C_{anchor}=\emptyset$}{
        $B\leftarrow E[i,:]$\;
        $\CorrectErrors(E[i,:])$\;
        \If(\tcp*[h]{disallow more than 2 bit corrections}){$\HammingWeight(E[i,:]-B)>2$}{
          $E[i,:]\leftarrow B$\;
        }
        \Else {
          $C\leftarrow C\cup E_r$\;
        }
      }
    }
    $R\leftarrow \emptyset$\;
  }
}
\caption{Modified iterative bounded-distance decoder.}
\label{alg:main_algorithm}
\end{algorithm}

\begin{algorithm}
\SetKwFunction{FindErrors}{FindErrors}
\SetKwFunction{CorrectErrors}{CorrectErrors}
\SetKwFunction{CorrectErasures}{CorrectErasures}
\SetKwFunction{False}{False}
\SetKwFunction{True}{True}
\SetKwFunction{And}{AND}
\SetKwFunction{Not}{NOT}
\SetKwFunction{HammingWeight}{HammingWeight}
\KwData{Encoded data $E\in \mathbb{F}_{2^8}^{n_r\times n_c}$ with $n_r$ rows and $n_c$ columns, with a horizontal $RS$ code distance $d_r=n_r-k_r+1$;}
\KwResult{In-place error correction of $E$.}
\tcc{Detecting column miscorrection}
$C\leftarrow \{1, 2, \ldots, n_c\}$\;
$R\leftarrow \emptyset$\;
\For{$i=1, 2, \ldots, n_r$}{
  $P\leftarrow \FindErrors(E[i,:])$\;
  \If {$P\neq \emptyset$} {
    $C\leftarrow C\cap P$\;
    $R=R\cup \{i\}$\;
  }
}
\If{$|C|=1~\And~|R|\geq d_r$}{
  \For{$i=1, 2, \ldots, n_r$}{
    $\CorrectErrors(E[i,:])$\;
  }  
}
\tcc{Detecting minimum stall pattern}
$C\leftarrow \emptyset$\;
\For{$j=1, 2, \ldots, n_c$} {
  \If{\Not $\CorrectErrors(E[:,j])$}{
    $C\leftarrow C\cup \{j\}$\;
  }
}
\If{$|C|\leq d_r-1$}{
  \For{$i=1, 2, \ldots, n_r$}{
    $\CorrectErasures(E[i,:], C)$\;
  }  
}
\caption{Postprocessing}
\label{alg:post_processing}
\end{algorithm}

With the ideas mentioned above, we are ready to formulate the main
algorithm \autoref{alg:main_algorithm} and the postprocessing
\autoref{alg:post_processing}. In addition to what we've discussed
above, \autoref{alg:main_algorithm} introduces explicit tracking of
modified rows/columns; it is also important that the algorithm starts
from a column iteration to reduce the number of miscorrections produced.

The algorithms use the following auxiliary functions for Reed--Solomon
codes:

\begin{itemize}
\tightlist
\item
  \(\FindErrors(\mathbf{y})\) --- returns the set of at most
  \(\left\lfloor\frac{n-k}{2}\right\rfloor\) error positions of
  \(\mathbf{y}\), or \(\emptyset\) if no errors are detected or
  positions cannot be reliably detected.
\item
  \(\CorrectErrors(\mathbf{y})\) --- in-place error correction of
  \(\mathbf{y}\); returns \(\True\) if the resulting input data is a
  valid codeword and \(\False\) otherwise.
\item
  \(\CorrectErasures(\mathbf{y}, C)\) --- corrects up to \(d-1\) errors
  at the provided positions \(C\) of the \(RS\) code.
\item
  \(\HammingWeight(\mathbf{y})\) --- returns the Hamming weight of the
  binary representation of \(\mathbf{y}\), e.g.~the number of \(1\) bits
  in \(\mathbf{y}\).
\end{itemize}

\subsection{Analysis of the modified
IBDD}\label{analysis-of-the-modified-ibdd}

It is well-known from the literature (see
\citeproc{ref-product_code_lower_bound}{{[}23{]}}) that the stall
pattern for the iterative decoder of an
\(RS(n_1, k_1)\times RS(n_2, k_2)\) product code is a submatrix of size
\(s_1\times s_2\) such that
\(s_1\geq t_1=\left\lfloor\frac{n_1-k_1}{2}\right\rfloor+1, s_2\geq t_2=\left\lfloor\frac{n_2-k_2}{2}\right\rfloor+1\).
The maximal by inclusion pattern is sometimes referred to as
\((t_1, t_2)\)-core. The following lemma is folklore.

\begin{lemma}[]\protect\hypertarget{lem-infinite-decoder}{}\label{lem-infinite-decoder}

Miscorrection-free IBDD with unlimited iterations

\begin{itemize}
\tightlist
\item
  Stops in finite time
\item
  Stops with either no errors or the \((t_1, t_2)\)-core
\end{itemize}

\end{lemma}

\begin{proof}
The first claim is standard: the iterative decoder stops when an
iteration does not introduce changes. In a miscorrection-free decoding,
each change reduces the number of errors and thus the maximum number of
iterations is at most the number of initial errors.

The second claim is trickier. First, note that the \((t_1, t_2)\)-core
is immutable under the iterative decoder: the main property of the core
is that none of the errors can be corrected by the row and column
decoders since for both the error correction capacity is exceeded. Thus,
the core should remain after termination. Second, suppose that we ended
up with something other than the \((t_1, t_2)\)-core; then there should
be at least one row or column within the error correction capability,
and thus the decoder will not terminate until only the core is left.
\end{proof}

\begin{figure}

\begin{minipage}{0.50\linewidth}

\centering{

\includegraphics[width=0.8\linewidth,height=\textheight,keepaspectratio]{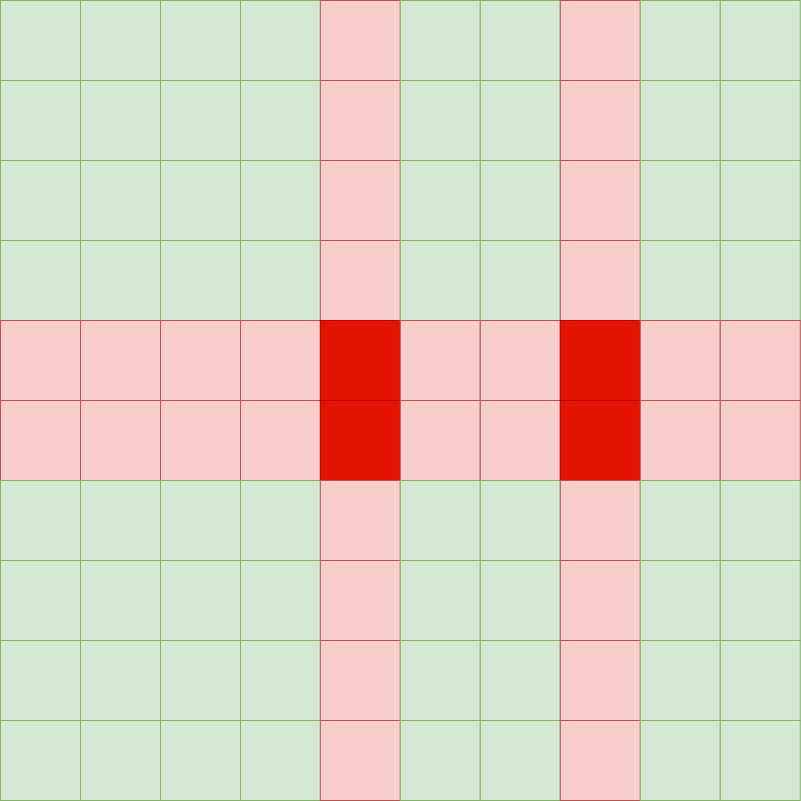}

}

\subcaption{\label{fig-min-matrix}Minimum stall pattern}

\end{minipage}%
\begin{minipage}{0.50\linewidth}

\centering{

\includegraphics[width=0.8\linewidth,height=\textheight,keepaspectratio]{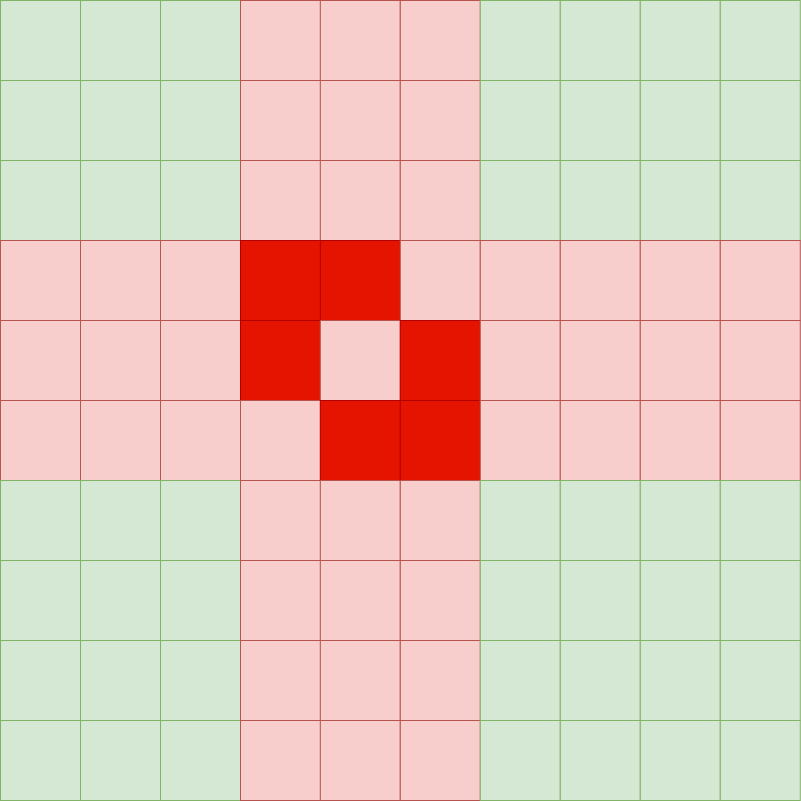}

}

\subcaption{\label{fig-second-min-matrix}Second minimum stall pattern}

\end{minipage}%

\caption{\label{fig-stall-patterns}On the left, the minimum stall
pattern for \(RS(10, 8)\times RS(10, 8)\) forming a \((2,2)\)-core where
\(4\) errors are on the intersection of two rows and two columns. On the
right, the second minimum pattern forming another \((2, 2)\)-core where
\(6\) errors are on the intersection of three rows and three columns so
that every column and every row contains \(2\) errors.}

\end{figure}%

For \autoref{alg:main_algorithm}, a similar analysis applies.

\begin{lemma}[]\protect\hypertarget{lem-our-algorithm}{}\label{lem-our-algorithm}

\autoref{alg:main_algorithm}

\begin{itemize}
\tightlist
\item
  Stops in finite time
\item
  Let \(E\) be the initial positions of errors and \(E'\) be the
  positions of the row miscorrections. Then after termination we are
  left with at most a \((t_1, t_2)\)-core formed from \(E\cup E'\).
\end{itemize}

\end{lemma}

\begin{proof}
After either successful correction or miscorrection of the column, it
cannot be further modified and thus the maximum number of iterations is
at most double the number of columns, i.e.~\(2n_2\).

The logic behind the second claim is the same as for the
miscorrection-free case.
\end{proof}

The important factor is that in practice \autoref{alg:ibdd} can enter a
loop when several bytes are modified in both column and row iterations
due to miscorrections, typically oscillating between two states until
the limit is reached. Introducing anchors for columns solves that issue.

Our last argument about \autoref{alg:main_algorithm} is empirical: in
practice we see that few miscorrections occur after our anchors and
binary image heuristics, and their proportion stays at around
\(\approx 7-10\%\) of initial errors; thus, the algorithm should perform
at a level of \(\approx 1.1\) raw BER of miscorrection-free IBDD.

\subsection{Analysis of
post-processing}\label{analysis-of-post-processing}

\subsubsection{\texorpdfstring{Stall patterns and
\((\alpha, \beta)\)-cores}{Stall patterns and (\textbackslash alpha, \textbackslash beta)-cores}}\label{stall-patterns-and-alpha-beta-cores}

Unfortunately, there is no known closed-form formula for either the
probability of existence of an \((\alpha, \beta)\)-core or the
expectation of its size/number of edges. We do, however, get accurate
estimates of the smallest such cores. Adapting the formula from
\citeproc{ref-product_codes_stall_patterns1}{{[}24{]}} gives us a
probability estimate for the minimum stall pattern
Fig.~\ref{fig-min-matrix}, i.e.~the probability of occurrence of a
\(t_1\times t_2\) submatrix with errors in each entry inside an
\(n_1\times n_2\) matrix as

\begin{equation}\phantomsection\label{eq-min-matrix}{
g_{\min}(p)=\frac{(b-1)p+1}{b}\frac{t_1t_2\binom{n_1}{t_1} \binom{n_2}{t_2} p_b^{t_1t_2}}{n_1n_2},
}\end{equation}

where \(p\) is the BER, \(b\) is the number of bits in a field element,
i.e.~\(8\) in our case, \(p_b=1-(1-p)^b\) is the probability of a field
element error.

A more accurate formula is given in
\citeproc{ref-product_codes_stall_patterns2}{{[}25{]}}; it includes many
higher-order stall patterns; an example is shown in
Fig.~\ref{fig-second-min-matrix}. The dominant part in our case is the
pattern with \(t_1+2\) rows and \(t_2+1\) columns with \(t_1(t_2+1)\)
errors:

\begin{itemize}
\tightlist
\item
  \(t_1-2t_2\) rows have \(t_2+1\) errors (all entries are errors)
\item
  \(2(t_2+1)\) rows each with \(t_2\) errors (all entries except one are
  errors)
\item
  Each column has exactly \(t_1\) errors
\end{itemize}

\begin{figure}

\centering{

\pandocbounded{\includegraphics[keepaspectratio]{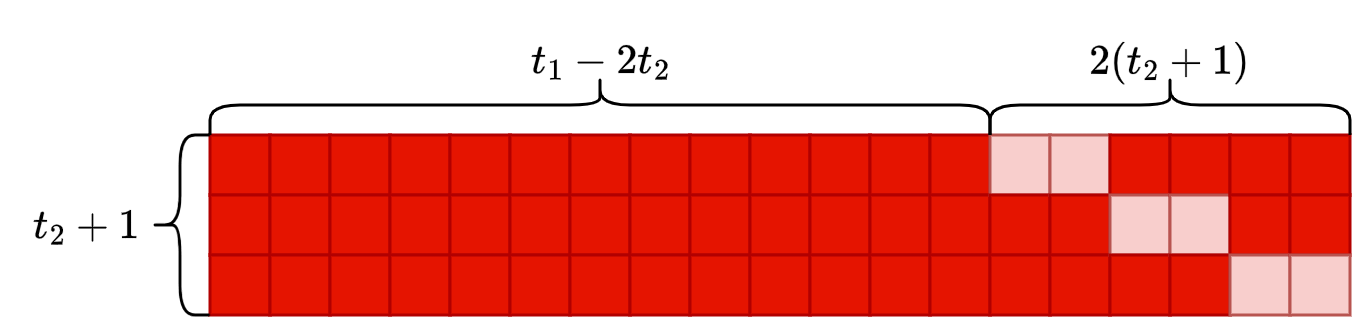}}

}

\caption{\label{fig-asymmetric-pattern}Non-trivial stall pattern,
transposed for visual alignment. For \(t_1=17\) and \(t_2=2\), this
results in a \(19\times 3\) submatrix with two parts: \(13\times 3\)
full of errors and \(6\times 3\) with two valid entries per column. The
bottom dashed line in Fig.~\ref{fig-product-code} is constructed via
this pattern for \(t_1=5, t_2=2\).}

\end{figure}%

This pattern is visualized in Fig.~\ref{fig-asymmetric-pattern} and the
corresponding error floor formula is
\begin{equation}\phantomsection\label{eq-second-matrix}{
\begin{aligned}
E_b(p)&=\frac{(b-1)p+1}{b} \\
N&=\frac{(t_1+2)!}{2^{t_2+1}(t_1-2t_2)!}\binom{n_1}{t_1+2} \binom{n_2}{t_2+1} \\
E&=\frac{t_1(t_2+1)}{n_1n_2}p_b^{t_1(t_2+1)}(1-p_b)^{2(t_2+1)} \\
\hat{g}_{\min}(p)&=E_b(p)NE,
\end{aligned}
}\end{equation}

where

\begin{itemize}
\tightlist
\item
  \(E_b(p)\) is the expected portion of bit errors in a \(b\)-length bit
  symbol provided that at least one bit error exists\footnote{\citeproc{ref-product_codes_stall_patterns1}{{[}24{]}}
    has a flaw in this coefficient; the denominator should be \(b\)
    rather than \(b^2\).}.
\item
  \(N\) is the number of stall patterns from
  \citeproc{ref-product_codes_stall_patterns2}{{[}25{]}} (page 75) of
  size \((d_1+1)(d_2+2)\).
\item
  \(E\) is the expected portion of errors multiplied by the probability
  of a particular \((t_1+2)(t_2+1)\) submatrix forming the pattern.
\end{itemize}

If the \emph{frame error rate} is needed, i.e.~the probability of having
an error after decoding, \(E_b(p)\) and \(\frac{t_1(t_2+1)}{n_1n_2}\)
should be omitted.

Under a raw BER of \(5\times 10^{-3}\), (\ref{eq-min-matrix}) gives an
estimate of \(\approx 10^{-22}\) and (\ref{eq-second-matrix}) an
estimate of \(\leq 10^{-34}\), which improves the lower bound
significantly.

\subsubsection{\texorpdfstring{Miscorrections of
\(RS(256, 224)\)}{Miscorrections of RS(256, 224)}}\label{miscorrections-of-rs256-224}

Anchors guarantee that the \(RS(175, 173)\) miscorrections will be
corrected/detected by a column iteration unless a miscorrection of a
column code also occurs. The problem with that case is that it is too
rare to affect the simulation but not rare enough to be discarded. In
particular, \citeproc{ref-rs_miscorrection_classical}{{[}30{]}} gives a
simple upper bound of

\begin{equation}\phantomsection\label{eq-base-miscorrection-prob}{
p_{mis}=\frac{p_t}{t!}
}\end{equation}

for miscorrection probability \(p_{mis}\) with \(p_t\) being the
probability of getting at least \(t\) errors. Under a raw BER of
\(5\times 10^{-3}\) this gives an \(\approx 10^{-17}\) miscorrection
probability with \(t=17\). (\ref{eq-base-miscorrection-prob}) can be
improved a little bit, but the improvement is not significant.

After \autoref{alg:main_algorithm} is finished, one of the following may
occur:

\begin{enumerate}
\def\labelenumi{\arabic{enumi}.}
\tightlist
\item
  No errors left or a stall pattern is left
\item
  A single miscorrection by a column code and no other errors left
\item
  A single miscorrection by a column code and at least one column with
  \(\left\lfloor\frac{n_2-k_2}{2}\right\rfloor+1\) errors left matching
  the row positions of miscorrected bytes
\item
  A single miscorrection by a column code and a stall pattern in other
  columns
\item
  Two or more miscorrections by a column code occurred
\end{enumerate}

Scenario 1 is trivial; scenarios 4 and 5 both have low probability.
Scenario 2 is provably solved by \autoref{alg:post_processing}. Finally,
scenario 3 is the one that cannot be resolved by postprocessing; there
might be some other, more complex constructions involving a single
column miscorrection and a few other uncorrected columns forming a stall
pattern, but they are less probable. Adjusting
(\ref{eq-base-miscorrection-prob}) with \(17\) bit errors over \(33\)
bytes of some other column and a pair-column variance, we get

\begin{equation}\phantomsection\label{eq-miscorrection-stall-prob}{
\frac{1}{17!}p^{17}(1-p)^{2031}\binom{2048}{17}p^{17}(1-p)^{247}\binom{264}{17}\binom{175}{2}
}\end{equation}

gives \(1.89\times 10^{-25}\) with \(p=5.4\times 10^{-3}\).

\section{Empirical evaluations}\label{sec-evaluations}

We provide two kinds of evaluations to support our claims; both are
Monte-Carlo simulations. The first one is a straightforward evaluation
of bit errors under a BSC; the second one is importance sampling of
column miscorrections.

\subsection{General MC simulation}\label{general-mc-simulation}

We organize our experiments in the following manner: the simulation is
essentially a large number of \(RS(256, 224) \times RS(175, 173)\)
codewords with bit errors induced according to a BSC. We group our
experiments by the number of bit errors to get an empirical estimate of
\(\mathbb{E}(u~|~e=k)\) with post-decoding bit errors \(u\) and raw bit
errors \(e\). Then, the estimate is plugged into

\begin{equation}\phantomsection\label{eq-main-curve}{
\mathbb{E}(u~|~p)\leq\sum_{k=0}^\infty \mathbb{E}(u~|~e=k)\times P(e=k~|~p)+\mathbb{E}_{stall}(p)
}\end{equation}

with \(P(e=k~|~p)\) being a binomial coefficient and
\(\mathbb{E}_{stall}(p)\) constructed via (\ref{eq-min-matrix}) or
(\ref{eq-second-matrix}). Four algorithm variants are given:

\begin{itemize}
\tightlist
\item
  \autoref{alg:ibdd} with \(4\) half-iterations
\item
  \autoref{alg:ibdd} with no iteration limit
\item
  \autoref{alg:main_algorithm} + \autoref{alg:post_processing} with no
  iteration limit
\item
  Ideal IBDD that does not make miscorrections with no iteration limit
\end{itemize}

Our estimate accuracy is down to \(10^{-10}-10^{-12}\); below that we
use extrapolation.

\begin{figure}

\centering{

\pandocbounded{\includegraphics[keepaspectratio]{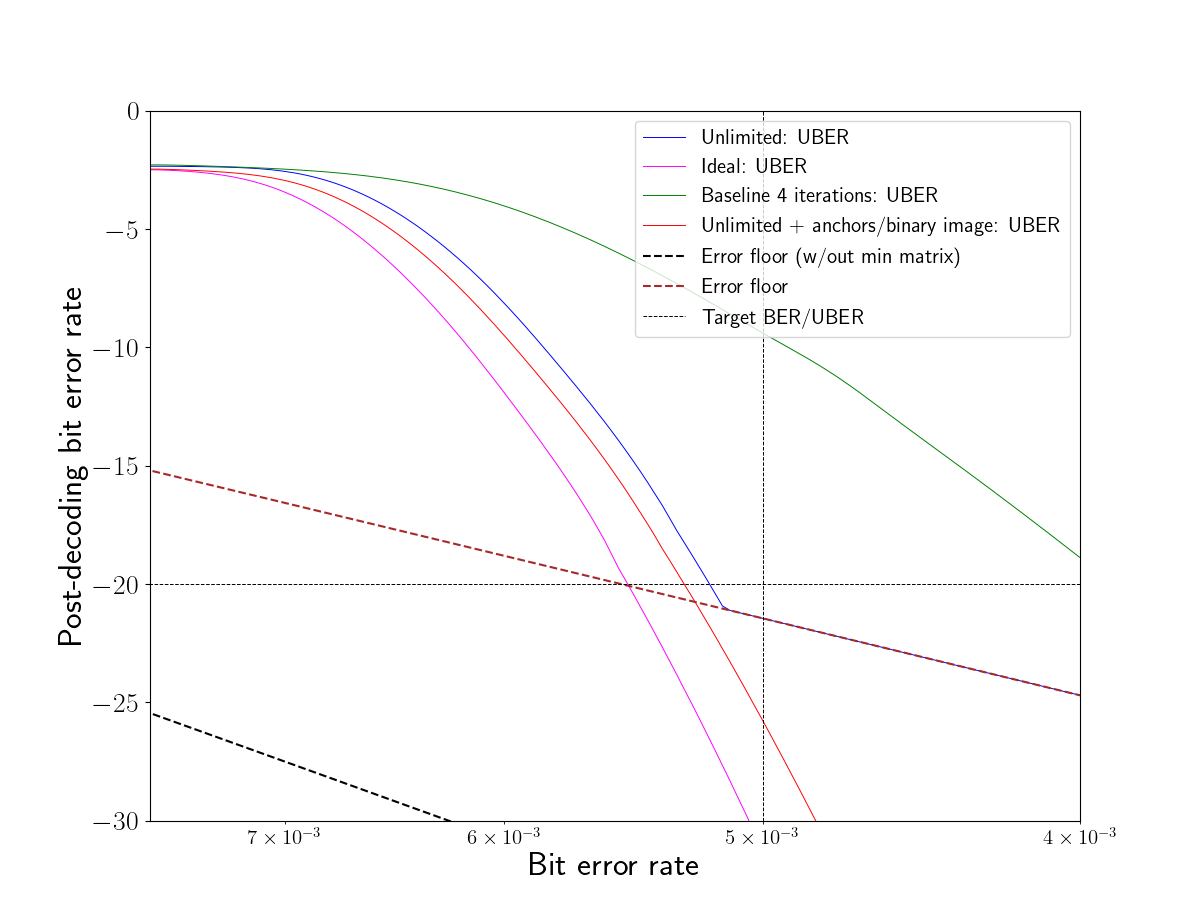}}

}

\caption{\label{fig-main-report-summary}Comparison among regular IBDD
with \(4\) and \(40\) iterations, our main algorithm and the ideal
decoder. The main algorithm achieves a post-decoding BER of \(10^{-20}\)
at \(\approx 5.4\times 10^{-3}\) raw BER.}

\end{figure}%

\begin{figure}

\centering{

\pandocbounded{\includegraphics[keepaspectratio]{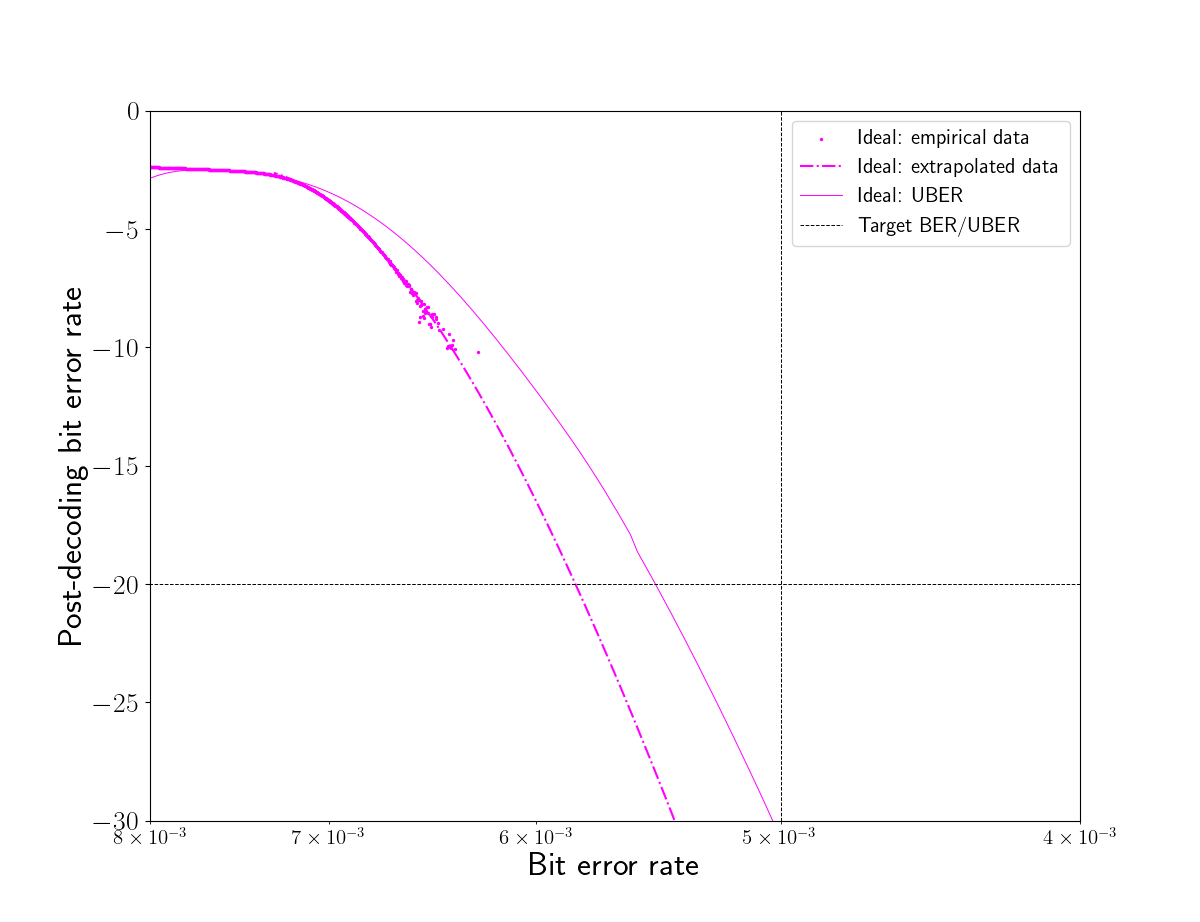}}

}

\caption{\label{fig-main-report-description}Post-decoding graph
description: \(\mathbb{E}(u~|~e=k)\) is estimated empirically down to
\(10^{-10}\) and extrapolated below, then plugged into
(\ref{eq-main-curve}); the resulting curve is the post-decoding BER.}

\end{figure}%

\begin{figure}

\centering{

\pandocbounded{\includegraphics[keepaspectratio]{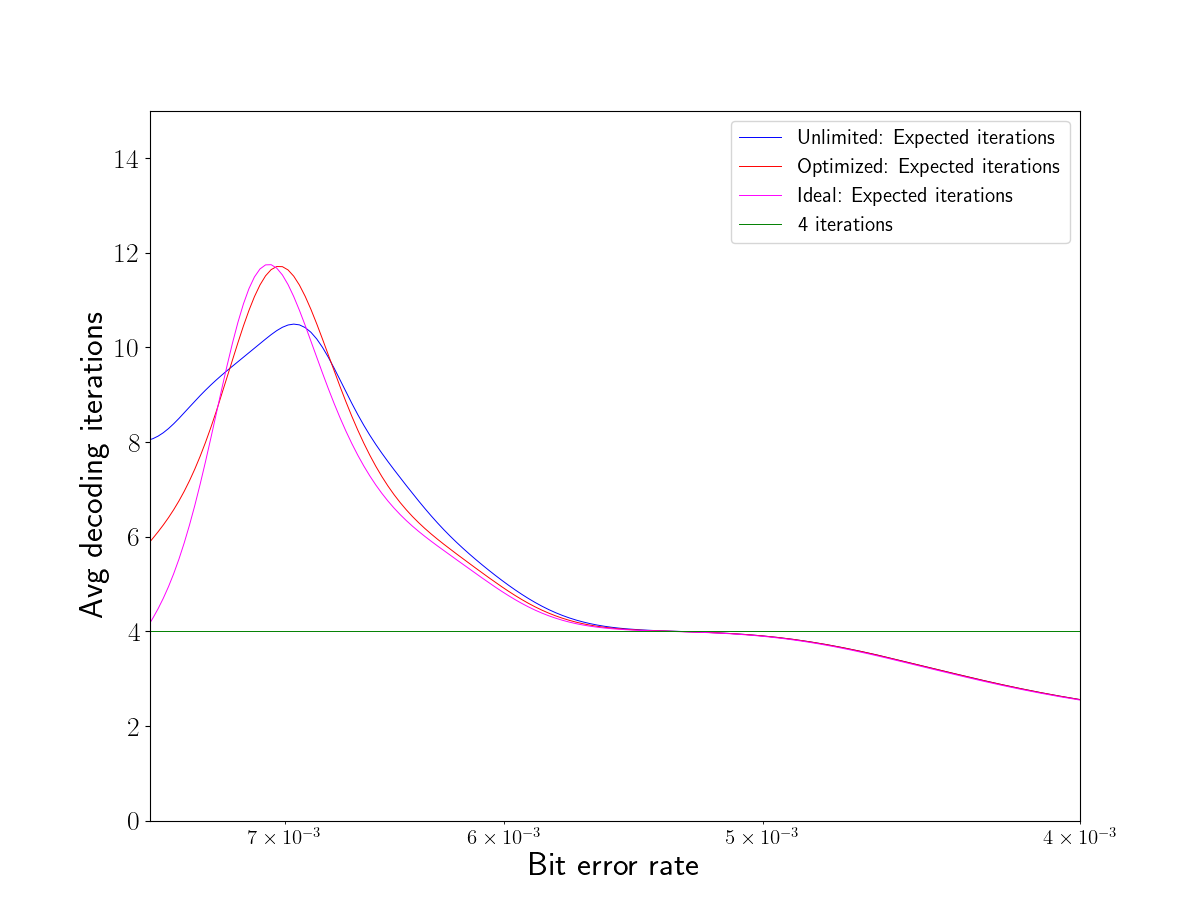}}

}

\caption{\label{fig-main-report-iterations}The average number of
iterations drops with the raw BER; for a raw BER of \(5\times 10^{-3}\)
the main algorithm performs fewer than \(4\) iterations on average.}

\end{figure}%

One of the most important observations is that while an unlimited
decoder performs fewer than \(4\) iterations on average for the raw BER
of \(5\times 10^{-3}\), it performs significantly better than the
decoder with \(4\) iterations. The explanation is simple: most of the
corruption patterns are recovered by the first \(2\)--\(3\) iterations,
and we only need to perform more in rare cases; in terms of reliability
this pays off.

\section{Discussion and conclusion}\label{sec-conclusion}

As a final note, we would like to give several thoughts on practical
usage of the provided algorithms. First of all, \autoref{alg:ibdd} is
commonly available on FPGAs, see
\citeproc{ref-smith2012_staircase}{{[}8{]}},
\citeproc{ref-furrer2021317}{{[}9{]}},
\citeproc{ref-arslan2022_lto9_uber}{{[}12{]}}. It is not clear whether
the anchor logic and binary image filters of
\autoref{alg:main_algorithm} are easily applicable. There is an
important structural property of the algorithm: the first two passes are
always full, and when applied under a reasonable raw BER the rest of the
iterations are performed only on a tiny subset of rows/columns.
Fig.~\ref{fig-main-report-summary} shows that these iterations are
important: there is a big difference between ``performing \(4\)
iterations'' and ``performing as many iterations as needed, which on
average is \(4\)''. For this reason, we believe that a hybrid FPGA-CPU
architecture might be a good choice for the algorithm:

\begin{itemize}
\tightlist
\item
  The FPGA performs two full half-iterations starting from the columns.
  The column iteration is performed normally and additionally saves
  information on whether the column has no syndromes after an iteration.
  The row iteration stores whether the codeword is already
  correct/contains a single mistake, and the corrected byte position and
  value if applicable (\(2\) bytes per row should be enough for all of
  that). This is then pipelined into the CPU.
\item
  The CPU performs anchor and binary image filtering and proceeds with
  the rest of the main loop; passing column information between
  iterations is also important to avoid full row/column scans in
  postprocessing.
\end{itemize}

Additionally, we provide a full CPU implementation in
\citeproc{ref-malkovsky_galois}{{[}31{]}} based on LCH-RS codes
\citeproc{ref-lin_chung_han_fft}{{[}32{]}} that achieves a single-thread
decoding throughput of \(120\) megabytes/s via AVX2 and \(200\)
megabytes/s via GFNI \citeproc{ref-gfni}{{[}33{]}}.

\section*{References}\label{references}
\addcontentsline{toc}{section}{References}

\phantomsection\label{refs}

\ifCLASSOPTIONcaptionsoff
  \newpage
\fi



\pagebreak[3]
\end{document}